# Multi-core fiber enabled fading noise suppression in φ-OFDR based quantitative distributed vibration sensing

Yuxiang Feng,[1] Weilin Xie,[1,2,*] Yinxia Meng,[1] Jiang Yang,[1] Qiang Yang,[1] Yan Ren,[1] Tianwai Bo,[1] Zhongwei Tan,[1] Wei Wei,[1,2] and Yi Dong[1,2]

[1]Key Laboratory of Photonics Information Technology, Ministry of Industry and Information Technology, School of Optics and Photonics, Beijing Institute of Technology, No. 5, South Street, Zhongguancun, Haidian District, Beijing 100081, P. R. China
[2]Yangtze Delta Region Academy of Beijing Institute of Technology, Jiaxing 314011, P. R. China
*Corresponding author: wlxie@bit.edu.cn

**Coherent fading has been regarded as a critical issue in phase-sensitive optical frequency domain reflectometry (φ-OFDR) based distributed fiber-optic sensing. Here, we report on an approach for fading noise suppression in φ-OFDR with multi-core fiber. By exploiting the independent nature of the randomness in the distribution of reflective index in each of the cores, the drastic phase fluctuations due to the fading phenomina can be effectively alleviated by applying weighted vectorial averaging for the Rayleigh backscattering traces from each of the cores with distinct fading distributions. With the consistent linear response with respect to external excitation of interest for each of the cores, demonstration for the propsoed φ-OFDR with a commercial seven-core fiber has achieved highly sensitive quantitative distributed vibration sensing with ~2.2 nm length precision and 2 cm sensing resolution along the 500 m fiber, corresponding to a range resolution factor as high as ~4×10$^{-5}$. Featuring long distance, high sensitivity, high resolution, and fading robustness, this approach has shown promising potentials in various sensing techniques for a wide range of practical scenarios.**

Past few decades have witnessed the rapid development of Rayleigh backscattering (RBS) based distributed fiber-optic sensing (DFOS) in fields ranging from industry and civil safety monitoring [1], defense and military [2] to scientific research [3]. In particular, owing to the linear response between the RBS phase and external changes of interest along the optical fibers [4], interrogation in terms of phase-sensitive optical frequency domain reflectometry (φ-OFDR) [5] that specifically features high sensitivity, resolution, and low processing bandwidth, has been regarded as a powerful tool in the implementation of DFOS [6] for quantitative vibration, acoustic, and dynamic strain measurement.

In general, to ensure a highly sensitive phase demodulation, coherent detection is usually applied, which nonetheless, leads to the problematic coherent Rayleigh fading. Due to the randomness in the distribution of the refractive index along the fiber, it would potentially result in points with a dramatically low signal-to-noise ratio (SNR) on the RBS trace due to the destructive interference [7] within the range of a single spatial resolution. This inevitably impairs the phase continuity, thus hampering the phase extraction during the demodulation.

As long been studied, this issue can be kind of mitigated relying on the averaging-based signal synthesis using certain statistically independent RBS measurements with their distinct fading phenomena [8]. In this context, the frequency-dependent nature of the RBS [9] has been exploited and has become popular owing to its practical feasibility that allows directly altering the phase relation, thus the fading distribution. It has been implemented in terms of frequency shift averaging by means of such as tunable laser, external modulation, and frequency comb [10, 11], with nevertheless, increased complexities due to the required large frequency tuning range. Alternatively, frequency-division multiplexing by internally splitting a linear frequency-swept probe allows to offer a series of frequency-separated RBS measurements [12, 13], enabling fading noise reduction, however, at the cost of a sacrificed resolution.

Inspired by the recent advent in space-division multiplexing (SDM) techniques in the telecom community, multi-mode [14] and few-mode fibers [15] that allow acquiring individual RBS profiles simultaneously from different spatial modes have been exploited for such averaging-based fading noise suppression. The crosstalks due to the non-negligible mode coupling, as a common issue in mode-division multiplexing, is still annoying and would probably degrade the sensing performance.

To this end, we present in this letter an efficient approach for fading noise suppression in φ-OFDR by exploiting the independence nature for the randomness in the reflective index along the cores of multi-core fibers (MCFs). The high isolation, i.e. low cross-talk, between different cores has allowed for effective fading noise suppression relying on the weighted vectorial averaging utilizing RBS traces from each of the cores with their different fading distributions. With the consistent response with respect to the external excitation of interest, distributed sensing is readily permitted for various physical quantities. The demonstrated φ-OFDR consisting of a commercial seven-core fiber (SCF) has achieved long-distance highly sensitive quantitative distributed vibration sensing with a significant range resolution factor, confirming the capability for practical fading noise mitigation.

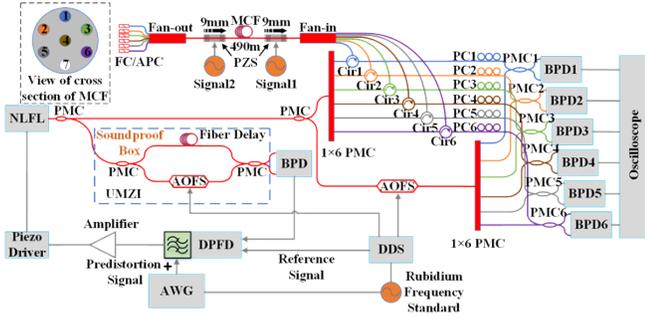

Fig.1 The proposed φ-OFDR with MCF acting as sensing medium. NLFL: narrow-linewidth fiber laser, PMC: polarization-maintaining coupler, AOFS: acousto-optic frequency shifter, BPD: balanced photodetector, DPFD: digital phase-frequency discriminator, DDS: direct digital frequency synthesizer, AWG: arbitrary waveform generator, Cir: circulator, PC: polarization controller, PZS: piezo stack.

In the proposed system, as schematically shown in Fig. 1, a narrow linewidth fiber laser with ~5 kHz linewidth acts as the probing source with its output split into the phase-locking and measurement parts, respectively. The former incidents into an unbalanced Mach-Zehnder interferometer (UMZI) with 2 km fiber delay followed by a digital phase frequency discriminator, with an arbitrary waveform generator, loop filter, and piezo driver for the phase error extraction, ramp generation, signal processing, and feedback control, respectively. These all together consist of an optical phase-locked loop to yield the required linearized frequency-swept optical probe as functionally elaborated in a similar setup [16]. The optimization for enhancing the coherence of the probe has been applied [17] to the UMZI before being placed in a home-made wooden soundproof box for isolating the influences of environmental perturbations. With the same laser, an optical probe of an 8 GHz sweep range and 160 GHz/s sweep rate is obtained with a repeating period of 100 ms, including the settle and recovery time for the laser.

Meanwhile, two 1×N polarization maintaining couplers (PMCs) with the associated circulators and 2×2 PMCs are employed to constitute a set of identical interferometers to interrogate each of the cores in a MCF. Polarization controllers (PCs) are adopted for the compact polarization alignment that maximizes the interference efficiency [18].

Aiming at a practical demonstration with sufficient scalability, here a commercially available SCF with a length of ~500 m has been adopted in connection with 1×6 PMCs for the implementation of six identical interrogators. Two piezo stacks (PZSs) with 9 mm length independently driven by function generators are, respectively, installed at ~1 m and ~490 m of the sensing MCF to simulate perturbations of interest such as axial vibrations. A multi-channel oscilloscope is used to acquire the output from the balanced photodetectors.

The result obtained from single-core, which essentially represents that using single-mode fiber (SMF), is shown in Fig. 2. It is observed that as a result of the coherent fading, significant intensity fluctuations can be found on the trace all along the fiber with even lower intensities than the noise floor as zoomed-in at the proximal and distal ends respectively in Fig. 2(b) and 2(c). Such distribution of weak SNR points has inevitably brought about the drastic variations in the phase spectrum of the trace, leading to the sharp phase jumps over 2π after the phase differential as seen accordingly for instance, in Fig. 2(d) and 2(e) at above distances. The resulting phase errors regardless of the distance make it difficult to distinguish the vibration along the fiber via phase demodulation.

With the consistent parameters but distinct spatial distributions of the refractive index, it allows for obtaining RBS traces with independent fading distributions from each of the cores. Exploiting the discrepancies in the amplitude and phase at the fading locations among these traces by means of a weighted vectorial averaging [19], where the rotation and the average between the RBS trace from different cores are weighted by their respective amplitudes, efficient suppression of the fading induced deterioration can be expected.

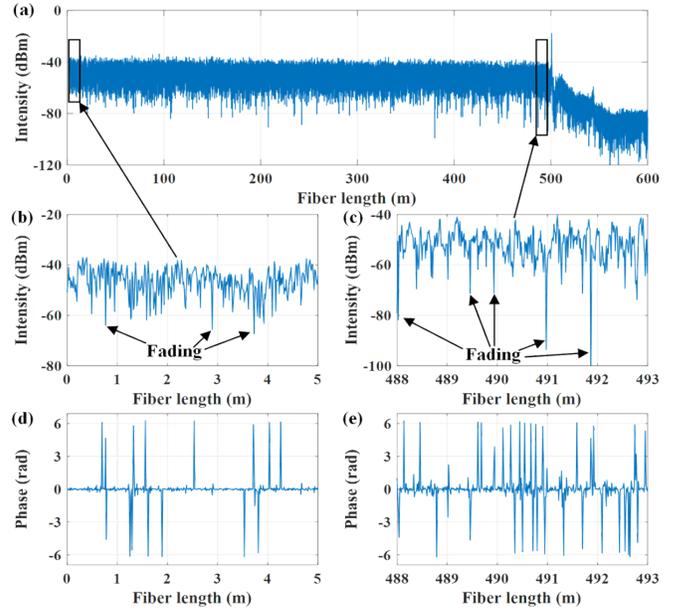

Fig.2 (a) Intensity trace obtained by φ-OFDR for a 500 m SCF in single-core case, which mimics that for SMF. Zoomed-in at sections from (b) 0 to 5 m and (c) 488 to 493 m with their corresponding differential phase in (d) and (e), respectively.

Such a process in the proposed system is illustrated in Fig. 3(a), where an obvious improvement in the variance of the RBS intensity is directly observed as the reduction in the thickness of the RBS trace in case of the average of 2 and 6 cores when compared with that from the single-core case, testifying for the effective fading noise suppression. Concerning a noise floor of about -80 dBm, as indicated in the measurement, it is observed that the lower bound for the intensity SNR of the fading points has been increased by at least ~7 dB after the average of 2 cores, while that in case of 6 cores has achieved ~26 dB. Fig. 3(c) summarizes the effect of fading suppression quantitatively in terms of the standard deviation (STD) of the intensity of the traces when different numbers of cores participate in the averaging. Compared with the single-core case, i.e. measurement based on SMF, the variance that corresponds to the impact of fading noise has been significantly suppressed by nearly four times with the averaging of 6 cores. The exhibited exponential relation agrees well with the tendency in previous theoretical studies [20].

The obvious fading noise suppression leads to remarkable reductions in the occurrence of 2π jump-overs after the phase differential along the trace as compared in Fig. 3(b) for the corresponding cases in Fig. 3(a). The dramatical phase leaps along the whole fiber in single-core case are effectively eliminated with the act of averaging using multiple cores, while a more smooth differential phase can be further seen in the case of 6-core average. Worth noting that a delicate trend where the variance slightly deteriorates along with the increase of distance can be identified. This is quantitatively explored by calculating the variance of the differential phase at every 50 m along the SCF as depicted in Fig. 3(d). The variation in the single-core case almost exceeds 1.5 rad$^2$ along the fiber, where the tendency is quite obscured due to the fading induced fierce phase jumps and is thus eliminated for the sake of clarity. Besides, the rest cases reveal the fact that the effect of fading noise suppression gets gradually diminished with the increase of the distance. This is probably attributed to the degradation due to the laser phase noise in particular at long distances. Assuming a certain precision that often requires a phase variance to be sufficiently small, e.g. less than 0.02 rad$^2$, it is then verified that a simple 2-core averaging would allow for the

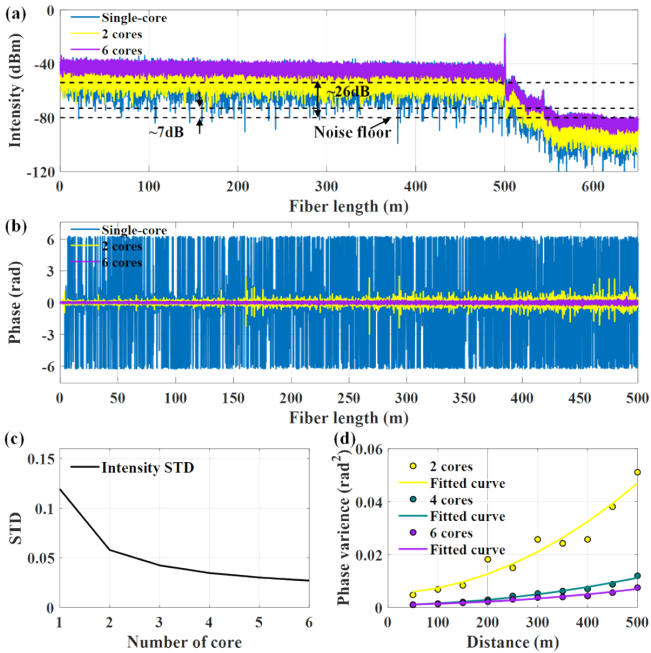

Fig.3 (a) Intensity traces obtained in single-, 2-, and 6-core cases, with (b) their corresponding differential phases. (c) The intensity STD with respect to the number of cores participated in averaging; (d) Evolution for the variances of the differential phase at every 50 m along the SCF.

measurement up to ~300 m, while a much longer range can be feasible by using more cores though limited according to both the fading noise and the phase noise of the laser.

Equivalent phase modulation caused by the external time-varying disturbances during sensing period is another source of randomization. Taking into account both the variations in time and distance, the fading effect is statistically analyzed in terms of its probability distribution with respect to the normalized intensity as given in Fig. 4(a) using the traces captured in a total 1000 measurements (duration of 100 s). An evident tendency inferred by the evolution of the intensity histogram reveals the concentration process for the intensity distribution. The wide and flat pedestal implying a large variation in single-core case has been clearly improved in a manner that a greater number of RBS becomes more centralized with increased intensity, namely a higher SNR, with the increase of the number of cores participating in the averaging. The most significant probability is found shifted to the location corresponding to ~10 dB SNR improvement with a narrowed distribution and almost twice the peak probability in the 6-core case. In addition, the corresponding cumulative distribution function (CDF) is derived, where if considering a -20 dB (i.e. 0.1) threshold, over 24% of the RBS obtained in single-core case could hardly beat this level. While in simply 2-core case, the amount drops rapidly to ~3% and even to ~0.03% in the case of averaging using 6 cores. Such statistical results strongly confirm the effectiveness of fading noise suppression with the averaging relying on the use of MCF.

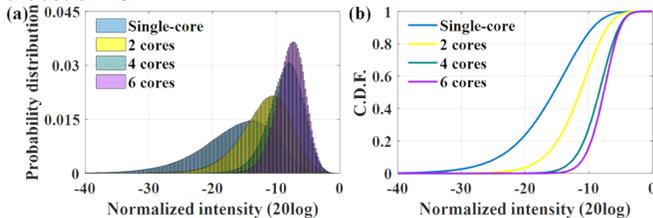

Fig.4 (a) Probability distributions for the normalized RBS intensities and (b) the corresponding CDF calculated using the results from 1000 φ-OFDR traces in single-, 2-, 4-, and 6-core cases.

With the capability of fading noise suppression for the proposed φ-OFDR, quantitative distributed vibration sensing is investigated via using the two PZSs. In the absence of multi-core averaging, the fading induced phase fluctuations dominate, hindering the demodulation for the actual vibration in either near or far end, as depicted in Fig. 5(a) and 5(b), respectively. Conversely, the obvious reduction in the variation of the differential phase in case of 6-core averaging has made it possible to accurately measure not only the location but also the frequency and amplitude of vibrations. Two vibration events at the proximal and distal ends can be clearly identified, simultaneously, as compared in Fig. 5(c) and 5(d) at ~1.46 m and ~490.14 m, respectively. The spatial resolution of 2 cm is readily achieved taking into account both the 8 GHz sweep range and the penalty brought by the Hanning window during the Fourier analysis, corresponding to a high range resolution factor up to ~4×10$^{-5}$ (ratio between spatial resolution and distance) for φ-OFDR based DFOS. The vibration waveforms along the entire SCF in 2 s, with 20 traces concerning the 100 ms measurement period in the case of 6-core averaging, are presented in a time-distance view as shown in Fig. 5(e), where the vibration resolution, i.e. the differential distance, is set the same as the spatial resolution. The locations of the time-varying changes are in line with the applied vibrations.

The differential phases that directly reflect the external vibrations are quantitatively characterized when the PZSs are independently driven with the same amplitude and frequency, as shown in Fig. 6(a) and 6(b), respectively. Even in a very short distance, it seems difficult to achieve a reliable measurement for single-core case, which is not surprising since coherent fading hardly depends on the optical path difference. While the comparisons with the 6-core case exhibit a high similarity with respect to the original driving signals, confirming the substantial improvement all along the fiber. Via spectrum analysis, we are able to extract the sensitivity for averaging different numbers of cores at both distances in terms of the vibration-induced length variation $\delta L$ as can be directly obtained by the minimum measurable phase changes $\Delta \varphi$ as $\delta L = \lambda \cdot \Delta \varphi /[4\pi(n + C_\varepsilon)]$, where $\lambda$, $n$, and $C_\varepsilon$ are the operation wavelength, refractive index, and strain coefficient [4], respectively. The SNR can also be derived in a similar manner. Obviously that with the increase of the number of averaging cores, the sensitivity quickly increases, reaching up to ~0.7 nm and ~2.2 nm in the 6-core case at the short and long distances, respectively. It is not difficult to find that laser phase noise induced phase variation plays an important role, especially in long distances, where the improvement in the sensitivity has somewhat deteriorated as implied in 2- and 3-core cases in Fig. 6(d). Similar trends can also be found for the SNR, where in long distances it is slightly lower than that in a short distances. These constitute a cross-verification for that unveiled in Fig. 3, verifying the effective fading noise suppression with the averaging using MCF in φ-OFDR based distributed sensing.

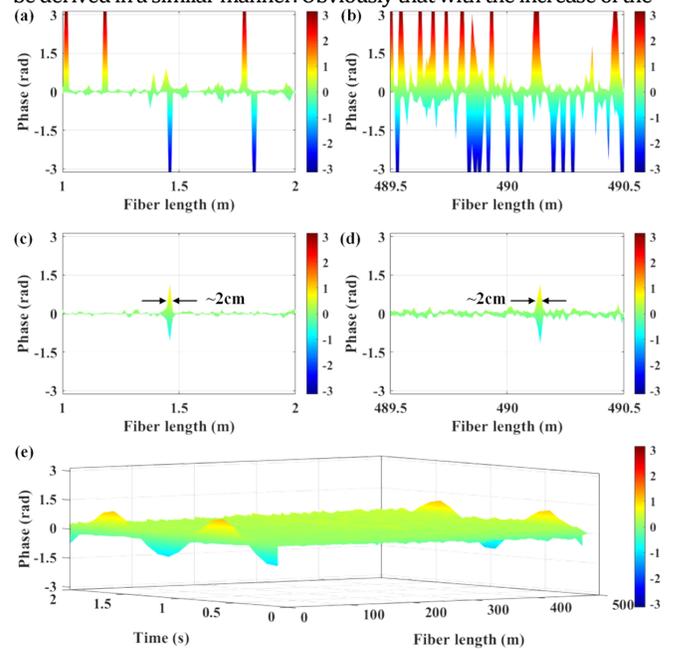

Fig.5 Detection for two independent vibration events in (a) and (b) for single- and (c) and (d) for 6-core cases, respectively at the proximal and distal ends of the SCF together with their (e) time-distance evolutions obtained with the traces from φ-OFDR measurement during 2 s.

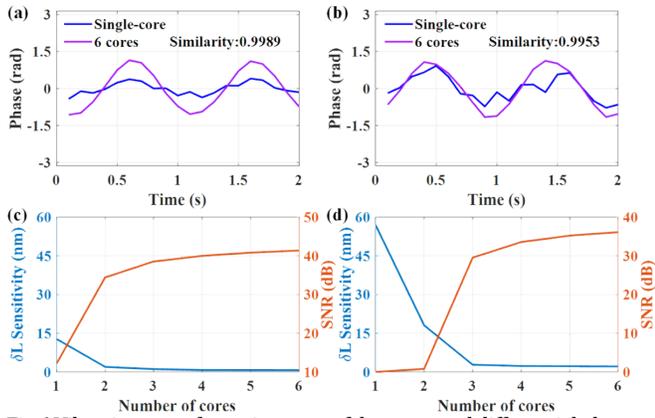

Fig.6 Vibration waveforms in terms of the extracted differential phase at (a) 1.46 and (b) 490.14 m, respectively, with the same but independent excitations in case of averaging with single and 6 cores; (c) and (d) are the sensitivity for the length variation and the SNR at the corresponding locations with respect to the number of cores involved in averaging.

In conclusion, by exploiting the distinct intrinsic randomnesses in the distribution of the refractive index together with the consistent linear relation to the external excitations for each of the cores in MCF, efficient fading noise suppression in φ-OFDR that enables long-distance and high resolution quantitative distributed vibration sensing has been verified by weighted vectorial averaging using RBS traces with distinct fading phenomena from each of the cores. A high sensitivity with ~2.2 nm length precision and 2 cm sensing resolution has been achieved along a 500 m range in an experimental demonstration using commercial SCF, corresponding to a range resolution factor as high as ~$4\times10^{-5}$. The potential exhibited by this approach offers practical feasibilities not only for the sensing of various physical quantities but also in different sensing techniques except for φ-OFDR.

**Funding.** National Natural Science Foundation of China (NSFC) (61827807, 61805014).

**Disclosure.** The authors declare no conflicts of interest.

**Data Availability.** Data underlying the results presented in this paper are not publicly available at this time but may be obtained from the authors upon reasonable request.

### References

1. M. Froggatt, and J. Moore, Appl. Opt. **37**, 1735 (1998).
2. J. Tejedor, H. F. Martins, D. Piote, J. Macias-Guarasa, J. Pastor-Graells, S. Martin-Lopez, P. C. Guillén, F. De Smet, W. Postvoll, and M. González-Herráez, J. Lightw. Technol. **34**, 4445 (2016).
3. P. Jousset, T. Reinsch, T. Ryberg, H. Blanck, A. Clarke, R. Aghayev, G. P. Hersir, J. Henninges, M. Weber, and C. M. Krawczyk, Nat. Commun. **9**, 2509 (2018).
4. Y. Dong, X. Chen, E. Liu, C. Fu, H. Zhang, and Z. Lu, Appl. Opt. **55**, 7810 (2016).
5. D. Arbel, and A. Eyal, Opt. Express **22**, 8823 (2014).
6. J. Li, J. Gan, Z. Zhang, X. Heng, C. Yang, Q. Qian, S. Xu, and Z. Yang, Opt. Express **25**, 27913 (2017).
7. P. Healey, Electron. Lett. **20**, 30 (1984).
8. K. Shimizu, T. Horiguchi, and Y. Koyamada, J. Lightw. Technol. **10**, 982 (1992).
9. H. Izumita, S. Furukawa, Y. Koyamada, and I. Sankawa, IEEE Photon. Technol. Lett. **2**, 201 (1992).
10. J. P. Von der Weid, R. Passy, G. Mussi, and N. Gisin, J. Lightw. Technol. **15**, 1131 (1997).
11. Z. He, T. Kazama, Y. Koshikiya, X. Fan, F. Ito, and K. Hotate, Opt. Express **19**, B764 (2011).
12. D. Chen, Q. Liu, and Z. He, Opt. Express **25**, 8315 (2017).
13. J. Jiang, Z. Wang, Z. Wang, Z. Qiu, C. Liu, and Y. Rao, Opt. Lett. **46**, 685 (2021).
14. A. E. Alekseev, V. S. Vdovenko, B. G. Gorshkov. V. T. Potapov, and D. E. Simikin, Laser Phys. **26**, 095101 (2016).
15. Z. Zhao, H. Wu, J. Hu, K. Zhu, Y. Dang, Y. Yan, M. Tang, and C. Lu, Opt. Express **29**, 15452 (2021).
16. Y. Feng, W. Xie, Y. Meng, L. Zhang, Z. Liu, W. Wei, and Y. Dong, J. Lightw. Technol. **38**, 6227 (2020).
17. L. Zhang, W. Xie, Y. Feng, Y. Meng, Y. Bai, J. Yang, W. Wei, and Y. Dong, Opt. Express **30**, 1994 (2022).
18. Z. Ding, X. S. Yao, T. Liu, Y. Du, K. Liu, Q. Han, Z. Meng, J. Jiang, and H. Chen, IEEE Photon. Technol. Lett. **25**, 202 (2013).
19. D. Chen, Q. Liu, and Z. He, Opt. Express **25**, 8315 (2017).
20. J. Ohtsubo, and T. Asakura, Opt. Lett. **1**, 98 (1977).

## Full Reference

1. M. Froggatt, and J. Moore, "High-spatial-resolution distributed strain measurement in optical fiber with Rayleigh scatter," Appl. Opt. 37, 1735 (1998).
2. J. Tejedor, H. F. Martins, D. Piote, J. Macias-Guarasa, J. Pastor-Graells, S. Martin-Lopez, P. C. Guillén, F. De Smet, W. Postvoll, and M. González-Herráez, "Toward prevention of pipeline integrity threats using a smart fiber-optic surveillance system, " J. Lightw. Technol. 34, 4445 (2016).
3. P. Jousset, T. Reinsch, T. Ryberg, H. Blanck, A. Clarke, R. Aghayev, G. P. Hersir, J. Henninges, M. Weber, and C. M. Krawczyk, "Dynamic strain determination using fibre-optic cables allows imaging of seismological and structural features," Nat. Commun. 9, 2509 (2018).
4. Y. Dong, X. Chen, E. Liu, C. Fu, H. Zhang, and Z. Lu, "Quantitative measurement of dynamic nanostrain based on a phase-sensitive optical time domain reflectometer," Appl. Opt. 55, 7810 (2016).
5. D. Arbel, and A. Eyal, "Dynamic optical frequency domain reflectometry," Opt. Express 22, 8823 (2014).
6. J. Li, J. Gan, Z. Zhang, X. Heng, C. Yang, Q. Qian, S. Xu, and Z. Yang, "High spatial resolution distributed fiber strain sensor based on phase-OFDR," Opt. Express 25, 27913 (2017).
7. P. Healey, "Fading in heterodyne OTDR," Electron. Lett. 20, 30 (1984).
8. K. Shimizu, T. Horiguchi, and Y. Koyamada, "Characteristics and reduction of coherent fading noise in Rayleigh backscattering measurement for optical fibers and components," J. Lightw. Technol. 10, 982 (1992).
9. H. Izumita, S. Furukawa, Y. Koyamada, and I. Sankawa, "Fading Noise Reduction in Coherent OTDR," IEEE Photon. Technol. Lett. 2, 201 (1992).
10. J. P. Von der Weid, R. Passy, G. Mussi, and N. Gisin, "On the characterization of optical fiber network components with optical frequency domain reflectometry," J. Lightw. Technol. 15, 1131 (1997).
11. Z. He, T. Kazama, Y. Koshikiya, X. Fan, F. Ito, and K. Hotate, "High-reflectivity-resolution coherent optical frequency domain reflectometry using optical frequency comb source and tunable delay line," Opt. Express 19, B764 (2011).
12. D. Chen, Q. Liu, and Z. He, "Phase-detection distributed fiber-optic vibration sensor without fading-noise based on time-gated digital OFDR," Opt. Express 25, 8315 (2017).
13. J. Jiang, Z. Wang, Z. Wang, Z. Qiu, C. Liu, and Y. Rao, "Continuous chirped-wave phase-sensitive optical time domain reflectometry," Opt. Lett. 46, 685 (2021).
14. A. E. Alekseev, V. S. Vdovenko, B. G. Gorshkov. V. T. Potapov, and D. E. Simikin, "Fading reduction in a phase optical time-domain reflectometer with multimode sensitive fiber," Laser Phys. 26, 095101 (2016).
15. Z. Zhao, H. Wu, J. Hu, K. Zhu, Y. Dang, Y. Yan, M. Tang, and C. Lu, "Interference fading suppression in φ-OTDR using space-division multiplexed probes," Opt. Express 29, 15452 (2021).
16. Y. Feng, W. Xie, Y. Meng, L. Zhang, Z. Liu, W. Wei, and Y. Dong, "High-performance optical frequency-domain reflectometry based on high-order optical phase-locking-assisted chirp optimization," J. Lightw. Technol. 38, 6227 (2020).
17. L. Zhang, W. Xie, Y. Feng, Y. Meng, Y. Bai, J. Yang, W. Wei, and Y. Dong, "Modeling and optimization of an unbalanced delay interferometer based OPLL system," Opt. Express 30, 1994 (2022).
18. Z. Ding, X. S. Yao, T. Liu, Y. Du, K. Liu, Q. Han, Z. Meng, J. Jiang, and H. Chen, "Long measurement range OFDR beyond laser coherence length," IEEE Photon. Technol. Lett. 25, 202 (2013).
19. D. Chen, Q. Liu, and Z. He, "Phase-detection distributed fiber-optic vibration sensor without fading-noise based on time-gated digital OFDR," Opt. Express 25, 8315 (2017).
20. J. Ohtsubo, and T. Asakura, "Statistical properties of the sum of partially developed speckle patterns," Opt. Lett. 1, 98 (1977).